\documentclass[journal]{IEEEtran}

\usepackage{booktabs}
\usepackage{tabularx}

\usepackage{amssymb}
\usepackage{amsbsy}
\usepackage{amsmath}
\usepackage{bm}
\usepackage{verbatim}
\usepackage{cite}
\usepackage{mathrsfs}
\usepackage{amsfonts}
\usepackage{graphicx}
\usepackage[tight,footnotesize]{subfigure}
\usepackage[10pt]{moresize}
\usepackage{array}
\usepackage{color}
\usepackage{epsfig}
\usepackage{stfloats}
\usepackage{balance}

\usepackage{subfigure}

\usepackage[noend]{algpseudocode}

\usepackage{algorithmicx,algorithm}

\usepackage{cite}

\usepackage{setspace}
\usepackage{cases}

\usepackage{graphicx}
\usepackage{epstopdf}
\usepackage{multirow}
\usepackage{extarrows}
\newcommand{\subparagraph}{}
\usepackage{titlesec}

\titlespacing{\section}{0pt}{2ex plus .0ex minus .0ex}{1ex plus .0ex}
\titlespacing{\subsection}{0pt}{1.0ex plus .0ex minus .0ex}{1.0ex plus 0.0ex}
\titlespacing{\subsubsection}{0pt}{0.3ex plus .0ex minus .0ex}{0.3ex plus .0ex}
\usepackage{amsmath} 
\allowdisplaybreaks[4]

\ifCLASSINFOpdf

\else

\fi

\begin{document}

\title{GSM: A GNN-based Space-MIMO Framework for Direct-to-Cell Communications}
\author{{Sai~Xu,~\IEEEmembership{Member,~IEEE,}
Yanan~Du,~\IEEEmembership{Member,~IEEE,}
Gaojie~Chen,~\IEEEmembership{Senior~Member,~IEEE,}
and~Rahim~Tafazolli,~\IEEEmembership{Senior~Member,~IEEE}  \vspace{-0.2 cm}

\thanks{%
Sai~Xu, Gaojie~Chen, and Rahim~Tafazolli (e-mail: sai.xu@ieee.org, gaojie.chen@surrey.ac.uk, r.tafazolli@surrey.ac.uk)  are with the Institute for Communication Systems (ICS), 5GIC \& 6GIC, University of Surrey, Guildford, Surrey GU2 7XH, U.K. Yanan~Du (e-mail: yanan.du@sheffield.ac.uk) is with the Department of Electronic and Electrical Engineering, University of Sheffield, Sheffield, S1 4ET, U.K.
}
} 
}

\maketitle
\begin{abstract}
This paper proposes a graph neural network (GNN)-based space multiple-input multiple-output (MIMO) framework, named GSM, for direct-to-cell communications, aiming to achieve distributed coordinated beamforming for low Earth orbit (LEO) satellites. Firstly, a system model for LEO multi-satellite communications is established, where multiple LEO satellites collaborate to perform distributed beamforming and communicate with terrestrial user terminals coherently. Based on the system model, a weighted sum rate maximization problem is formulated. Secondly, a GNN-based method is developed to address the optimization problem. Particularly, the adopted neural network is composed of multiple identical GNNs, which are trained together and then deployed individually on each LEO satellite. Finally, the trained GNN is quantized and deployed on a field-programmable gate array (FPGA) to accelerate the inference by customizing the microarchitecture. Simulation results demonstrate that the proposed GNN scheme outperforms the benchmark ones including maximum ratio transmission, zero forcing and minimum mean square error. Furthermore, experimental results show that the FPGA-based accelerator achieves remarkably low inference latency, ranging from 3.863 to 5.883 ms under a 10-ns target clock period with 8-bit fixed-point data representation.
\end{abstract}
%
%
%
\IEEEpeerreviewmaketitle
%
\section{Introduction}
\IEEEPARstart{W}ITH a substantial reduction in satellite production and launch costs, direct-to-cell (D2C) or direct-to-Earth (DTE) communications have gradually become an active research field in both academia and industry~\cite{Tuzi2023Distributed}. Compared with terrestrial communication networks, low Earth orbit (LEO) satellite networks are capable of realizing global seamless coverage, flexibility, and strong resilience~\cite{Deng2021Ultra, Xu2023Enhancement, Xu2021Envisioning}. With these inherent advantages, LEO satellite networks provide an ideal communication way in many practical scenarios, such as Earth observation and mapping, maritime affairs, disaster rescue, etc~\cite{Heo2023MIMO}. However, satellite communications are vulnerable to large signal attenuation due to path loss, the Earth's atmosphere and rainfall effects~\cite{Baktur2022CubeSat}. Moreover, large Doppler shift is an intractable issue~\cite{Ali1998Doppler, Zuo2023OFDM}. As a result, current D2C communications are still insufficient to support many high-throughput and low-latency services.\par
Facing the challenge, one promising technique is to combine multiple LEO satellites into a distributed-antenna transmitter to collaboratively communicate with terrestrial user terminals (UTs), forming a space multiple-input multiple-output (Space-MIMO) system. In principle, Space-MIMO is similar to cell-free MIMO for terrestrial networks~\cite{Ngo2017CF}. As a new paradigm, how to take advantage of the Space-MIMO technology to achieve high throughput and low latency remains an open topic. In this context, this paper will focus on a key issue: how to implement coordinated beamforming for LEO multi-satellite communications in a lightweight and distributed manner.\par
\subsection{Motivations}
To enhance D2C communications, this paper proposes a graph neural network (GNN)-based Space-MIMO framework for distributed coordinated beamforming of multiple LEO satellites. This framework encompasses the Space-MIMO technology, a multi-GNN architecture, and a field-programmable gate array (FPGA)-based GNN accelerator. The main motivations are outlined as follows.\par
\subsubsection{Why Make Use of the Space-MIMO Technology?}
In space-to-ground communications, a single satellite is often limited to power budget. Additionally, the communications suffer much from signal fading, obstructions, transmission environment, etc. In contrast, a cluster of satellites can enhance both the transmit-antenna and receive-antenna gains, as well as the link reliability, by constituting a constellation to form a distributed antenna array. Owing to spatial diversity, a number of available satellites with different visual angles can significantly improve space-to-ground communications. \par
\subsubsection{Why Adopt the GNN-based Optimization Method?}\par
In downlink multi-user satellite-terrestrial communications, inter-user and inter-satellite interference is one of the most critical factors restricting the communication performance. GNN excels at capturing the interactions between different signals carrying diverse data streams, provided their transmission channels are represented as a graph. Using GNN, a mechanism for interference suppression can be learned. Additionally, GNN is scalable, facilitating the construction of a multi-GNN architecture capable of effectively managing inter-satellite interference and scheduling switches among LEO satellites over time within the LEO multi-satellite system.
\subsubsection{Why Implement GNN on FPGA?}\par
GNN often involves a substantial number of weight and bias parameters, resulting in significant computational overhead and latency during inference. However, lightweight and low-latency designs must be prioritized for LEO satellites, given their power consumption budget and high-speed movement. Owing to low latency and power consumption, FPGA represents an ideal computational platform for handling the GNN workload. Hence, it is appealing to customize an application-specific FPGA-based GNN accelerator.
\subsection{State-of-the-Art}
\subsubsection{Multi-Satellite Communications}
%
%
A cluster of satellites can synchronize to communicate with UTs akin to a virtual MIMO system. Before LEO multi-satellite communications, the concept similar to Space-MIMO has appeared in geostationary orbit (GEO) satellite communications. Li \textit{et al.}~\cite{Li2021Analysis, Li2021Capacity} proved that collaborative satellites yield the same effect as implementing MIMO and analyzed the communication capacity of uniform linear array antenna and uniform circular array antenna. Zhao \textit{et al.}~\cite{Zhao2019Multisatellite} investigated the random access technique and Feng \textit{et al.}~\cite{Feng2020satellite} studied the inter-satellite handover technique. In the aspect of LEO multi-satellite communications,  Abdelsadek~\textit{et al.}~\cite{Abdelsadek2021Future, Abdelsadek2022Distributed, Abdelsadek2023Broadband} has published several works related to Space-MIMO, including the maximization of sum data rate, the minimization of handover rate, and the superiority of distributed MIMO to collocated MIMO.
\subsubsection{GNN-based Optimization Method}
GNN is a type of deep neural network (DNN) designed to process graph-structured data. It operates by propagating and aggregating information through the features of nodes and edges, effectively capturing relationships and patterns within the graph. With the aid of GNN, Jiang \textit{et al.}~\cite{Jiang2021Learning} directly mapped received pilots and user locations to beamforming vectors.  Li \textit{et al.}~\cite{Li2024GNN} maximized the sum rate of multi-user downlink communications. Zhang \textit{et al.}~\cite{Zhang2022Learning} achieved high overall throughput by scheduling users and configuring intelligent reflecting surface. Mishra \textit{et al.}~\cite{Mishra2024Graph} optimized power control in partially connected cell-free massive MIMO. Xu \textit{et al.}~\cite{Xu2023Distributed} investigated multi-cell cluster-free non-orthogonal multiple access communications. Chen \textit{et al.}~\cite{Chen2024Distributed} studied intelligent reflecting surface enhanced cell-free MIMO networks.
\subsubsection{FPGA-based Deep Learning Accelerator}
With reconfigurability, high energy efficiency, and rapid development cycles, FPGA is a frequently used hardware platform for accelerating the inference process of DNN. By a neural network-oriented hardware design, FPGA can achieve high parallelism and eliminate extra logic. In the field of FPGA acceleration, two kinds of architectural paradigms are often considered. The first one is a non-pipeline CPU-like architecture~\cite{Genc2021Gemmini, Yu2019OPU, Yu2020Light-OPU}, which adopts a generic compute unit to accelerate all the layers of DNN and has a frequent off-chip memory access and a low resource utilization. The second one utilizes a complete pipeline architecture~\cite{Wei2018TGPA, Zhang2020DNNExplorer}, leveraging a strategy that processes layers sequentially to optimize resources. 
\subsection{Contributions}
So far, distributed coordinated beamforming has not been thoroughly investigated in Space-MIMO empowered D2C communications. Moreover, the use of GNN and its application-specific FPGA-based accelerator remain unexplored in LEO multi-satellite communications. In light of these considerations, this paper proposes a GNN-based Space-MIMO framework to facilitate D2C communications. Our main contributions are summarized below.
\begin{itemize}
\item{This paper establishes a GNN-based Space-MIMO framework, named GSM\footnote{``GSM" is a double entendre, referring to both ``GNN-based Space-MIMO Framework" and ``Global System for Mobile Communications".}, designed to achieve high-quality D2C communications through co-design of algorithm, software and hardware. Particularly, the entire framework comprises three components: a system model for D2C communications, a multi-GNN architecture tailored for the optimization problem, and an FPGA-based accelerator for each individual GNN.}

\item{A system model of LEO multi-satellite communications is built. In the model, multiple LEO satellites are employed to constitute a constellation, which can implement coordinated beamforming for downlink multi-user communications. Based on this model, the maximization problem of weighted sum rate (WSR) is formulated under the constraint of transmit power at each LEO satellite. } 
\item{A GNN-based method is developed to address the optimization problem. Specifically, the whole communication network is associated with a multi-GNN architecture, where all the GNNs have exactly the same structure and each corresponds to one LEO satellite. The entire multi-GNN network is trained centralizedly, while each GNN is deployed distributedly on each LEO satellite. Furthermore, the GNN model is scalable and generalizes well, which enables the distributed deployment and independent implement of each individual GNN. }
\item{The trained GNN model is implemented on FPGA. To reduce the latency and the power consumption of computation, an FPGA-based accelerator is designed and optimized to deploy the trained GNN network by exploring the hardware parallelism of GNN inference. The design of FPGA-based GNN accelerator conforms to the latency-sensitive requirements of LEO multi-satellite communications. Note that this customization is necessary as the adopted neural network deviates from conventional GNN architectures, making existing FPGA-based GNN accelerators incompatible.}
\item{The numerical simulations are performed to evaluate the achievable communication performance of the developed GNN method. Furthermore, the experimental results are presented to show the advantages of the FPGA-based accelerator in improving GNN inference. } 
\end{itemize}

\subsection{Organizations}
The rest of this paper is organized as follows. In Section II, the system model is built and  the optimization problem is formulated. In Section III, a GNN-based method is presented for the optimization problem. In Section IV, the FPGA-based accelerator is designed. In Section V, the numerical and experimental results are presented. In Section VI, the conclusions are drawn.\\
\section{System Model and Problem Formulation}
\begin{figure}
\centering
\includegraphics[width=2.8in]{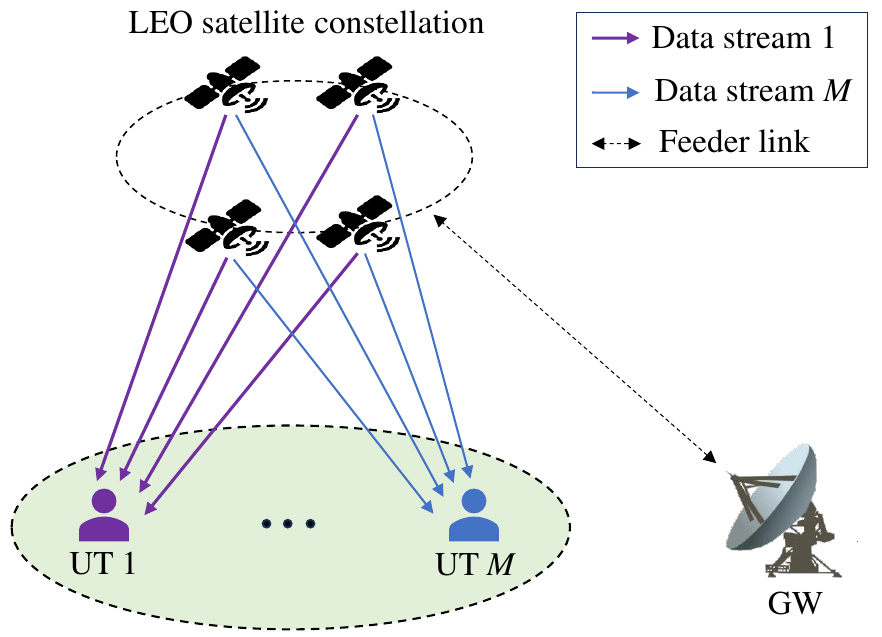}
\caption{A satellite-terrestrial Space-MIMO communication system, consisting of $K$ interconnected LEO satellites each equipped with $N$ antennas, $M$ single-antenna UTs, and a GW. }
\label{Fig1}
\end{figure}
Fig. \ref{Fig1} depicts a satellite-terrestrial Space-MIMO communication system, consisting of $K$ interconnected LEO satellites each equipped with $N$ antennas, $M$ single-antenna UTs, and a gateway (GW). Note that $N \geq M$ is set to ensure that the number of antennas of each LEO satellite is not less than that of the served UTs. In practice, the number of UTs far exceeds that of LEO satellite antennas, thus only a subset of UTs are allowed to access the satellite network at any given time slot in the narrowband transmission. Through the feeder link, the GW coordinates multiple LEO satellites to form a Space-MIMO transmitter, which can send $M$ data streams to the UTs.  Define $\mathcal{K} = \{1, 2, \cdots, K\}$ as the index set for all the LEO satellites and $\mathcal{M} = \{1, 2, \cdots, M\}$ as the index set for all the UTs.
\subsection{Channel Model}
In the considered system, all the satellite-terrestrial links are assumed to be block fading channels and are characterized as the classical Shadowed-Rician fading model incorporating the atmospheric effect~\cite{Xu2021Intelligent}. Mathematically, the satellite-terrestrial channel model is given by
\begin{equation}\label{eq1}
h = C_L \sqrt{ b(\varphi)} \tilde{h}. \nonumber
\end{equation}
\noindent In this expression, $C_L = {\lambda}/(4 \pi\sqrt{{ d_0^2 + d_h^2}})$ is the free-space path loss coefficient, where $\lambda$ is the carrier wavelength, $d_0$ denotes the height of LEO satellites, and $d_h$ represents the distance between the center of LEO satellite coverage area and the central beam. $\tilde{h}$ denotes the fading coefficient of satellite-terrestrial channels, which is given by
\begin{equation}\label{eq2}
\tilde{h} = A \text{exp}(j \psi) + Z \text{exp}(j \phi).  \nonumber
\end{equation}
 $\tilde{h}$ is made up of the scattering component $A \text{exp}(j \psi)$ and the line-of-sight (LOS) component $Z \text{exp}(j \phi)$, where  $\psi \in [0, \pi]$ and $\phi \in [0, \pi]$ represent the stationary random phase and the deterministic phase; $A$ and $Z$ obey Rayleigh distribution and Nakagami-\emph{m} distribution, respectively. $b(\varphi)$ denotes the beam gain factor, which is given by
\begin{equation}\label{eq3}
b(\varphi) = b_\text{max} \left( \frac{\text{J}_1(u)}{2u} + 36 \frac{\text{J}_3(u)}{u^3}\right)^2.  \nonumber
\end{equation}
\noindent $b(\varphi)$ depends on the UT's position and $b_\text{max}$ is the maximal value of satellite antenna gain. $\text{J}_1(\cdot)$ and $\text{J}_3(\cdot)$ denote the first-kind first-order and third-order Bessel functions, respectively. $u$ is given by
\begin{equation}\label{eq4}
u = 2.07123 \frac{\text{sin} \varphi }{\text{sin} (\varphi_\text{3dB}) },  \nonumber
\end{equation}
\noindent where $\varphi_\text{3dB}$ is the 3-dB angle and $\varphi$ denotes the angle between the two LOS paths from an LEO satellite to a UT and the beam center. \par
\subsection{Signal Model}
Let $\textbf{h}_{k,m} \in {\mathbb{C}^{N \times 1}}$, $\textbf{w}_{k,m} \in {\mathbb{C}^{N \times 1}}$, and $s_m$ denote the channel gain vector from the $k$-th LEO satellite to the $m$-th UT, the beamforming vector for the $m$-th data stream at the $k$-th LEO satellite, and the $m$-th data stream that is simultaneously sent by all the LEO satellites for the $m$-th UT, respectively, with $k \in \mathcal{K}$ and $m \in \mathcal{M}$. Assuming that all the LEO satellites can exchange their channel state information (CSI) with each other, the received signal at the $m$-th UT from all the LEO satellites is given by
\begin{align}
{y}_m  &=  \sum_{k=1}^K  \textbf{h}_{k,m}^H \sum_{i=1}^M \textbf{w}_{k,i} s_{i}  +  {n}_m \nonumber\\
 &= \sum_{k=1}^K  \textbf{h}_{k,m}^H  \textbf{w}_{k,m} s_m  + \sum_{i=1,i \neq m}^M  \sum_{k=1}^K  \textbf{h}_{k,m}^H \textbf{w}_{k,i} s_{i}  + {n}_m, \nonumber
\end{align}
where ${n}_m$ is white Gaussian random noise at the $m$-th UT with ${n}_m \sim \mathcal{CN}(0, \sigma^2 )$. Accordingly, the achievable communication rate at the $m$-th UT is given by
\begin{align}
R_m =  B \log_2\left ( 1 + \frac{ | \sum_{k=1}^K   \textbf{h}_{k,m}^H  \textbf{w}_{k,m} |^2}{ \sum_{i=1,i\neq m}^M  |\sum_{k=1}^K  \textbf{h}_{k,m}^H  \textbf{w}_{k,i} |^2 +\sigma^2} \right), \nonumber
\end{align}
where $B$ denotes the communication bandwidth.\par
\subsection{Problem Formulation}
\par The goal of this paper is to maximize the WSR by optimizing the beamforming vectors $\textbf{w}_{k,m}$, with $k \in \mathcal{K}$ and $m \in \mathcal{M}$. The optimization problem is formulated as
\begin{align*}
(\text{P1}) \quad \underset{\mathcal{W}} \max \quad &   \sum_{m = 1}^M \omega_{m} R_m, \\
s.t. \quad
 & \text{Tr} \left( \sum_{m = 1}^M \textbf{w}_{k,m} \textbf{w}_{k,m}^{H} \right) \leq P,~k \in \mathcal{K}, 
\end{align*}
where $\omega_{m}$ denotes the weighting factor of the $m$-th UT. $\mathcal{W}$ refers to the collection of $\textbf{w}_{k,m}$. $P$ is the transmit power budget of each LEO satellite.  
\section{GNN-based Optimization Method}
The optimization problem (P1) is difficult to tackle directly as the objective function takes the form of a weighted sum of logarithmic terms. Using Lagrangian dual transform, the objective function can be reformulated, followed by alternative optimization for solving the new problem. However, this approach involves high computational complexity and is prone to local optima. To seek an excellent solution to the non-convex problem (P1), this section will present a GNN-based optimization method. Firstly, a multi-GNN architecture is built for the system of LEO multi-satellite communications. Secondly, the structure of a single GNN is introduced. Thirdly, it is illustrated how to train and deploy the multi-GNN model. Lastly, the complexity analysis of the GNN-based optimization method is given.

\subsection{Multi-GNN Architecture and Graph Representation}
\begin{figure}
\centering
\includegraphics[width= 3.3 in]{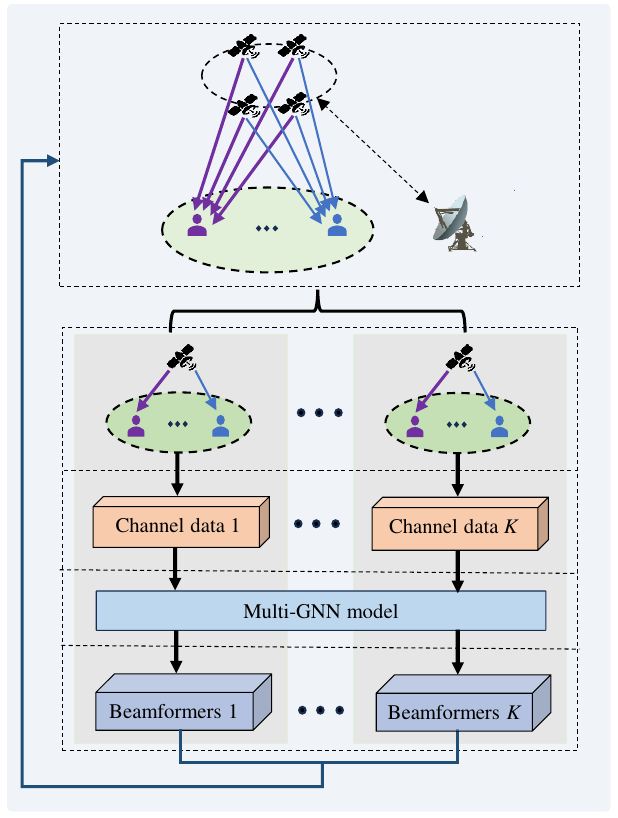}
\caption{The relationship between the proposed multi-GNN architecture and the system of LEO multi-satellite communications.}
\label{Fig2}
\end{figure}
Fig. \ref{Fig2} shows the relationship between the proposed multi-GNN architecture and the system of LEO multi-satellite communications. Depending on which LEO satellite the transmitted signal is from, the entire communication network can be partitioned into $K$ homogeneous local networks, each consisting of one LEO satellite and all the UTs associated with it.
Echoing the partition of the satellite-terrestrial network, the overall neural network is constructed as an integration of $K$ GNNs, each having the same structure. 
To capture the inner connection of all the components, a well-defined graph representation for each local network is an essential prerequisite. Here, the graph representation is composed of all the UT nodes and the edges that can characterize the correlation between any two UT nodes. \par
When the graph representation is fed into its paired GNN, the interaction among the UT nodes can be extracted by updating its graph representation vector multiple times. Through the entire multi-GNN model, the information related to the coexisting inter-satellite and intra-satellite interference is embedded in the output representation vector. Based on the system model and the problem (P1) in Section II, the input and output representation vectors for each GNN are set to the channel vectors of all the UTs and the optimization variables of the problem (P1), respectively. 
\subsection{Structure of GNN}
\begin{figure*}
\centering
\includegraphics[width= 6.8in]{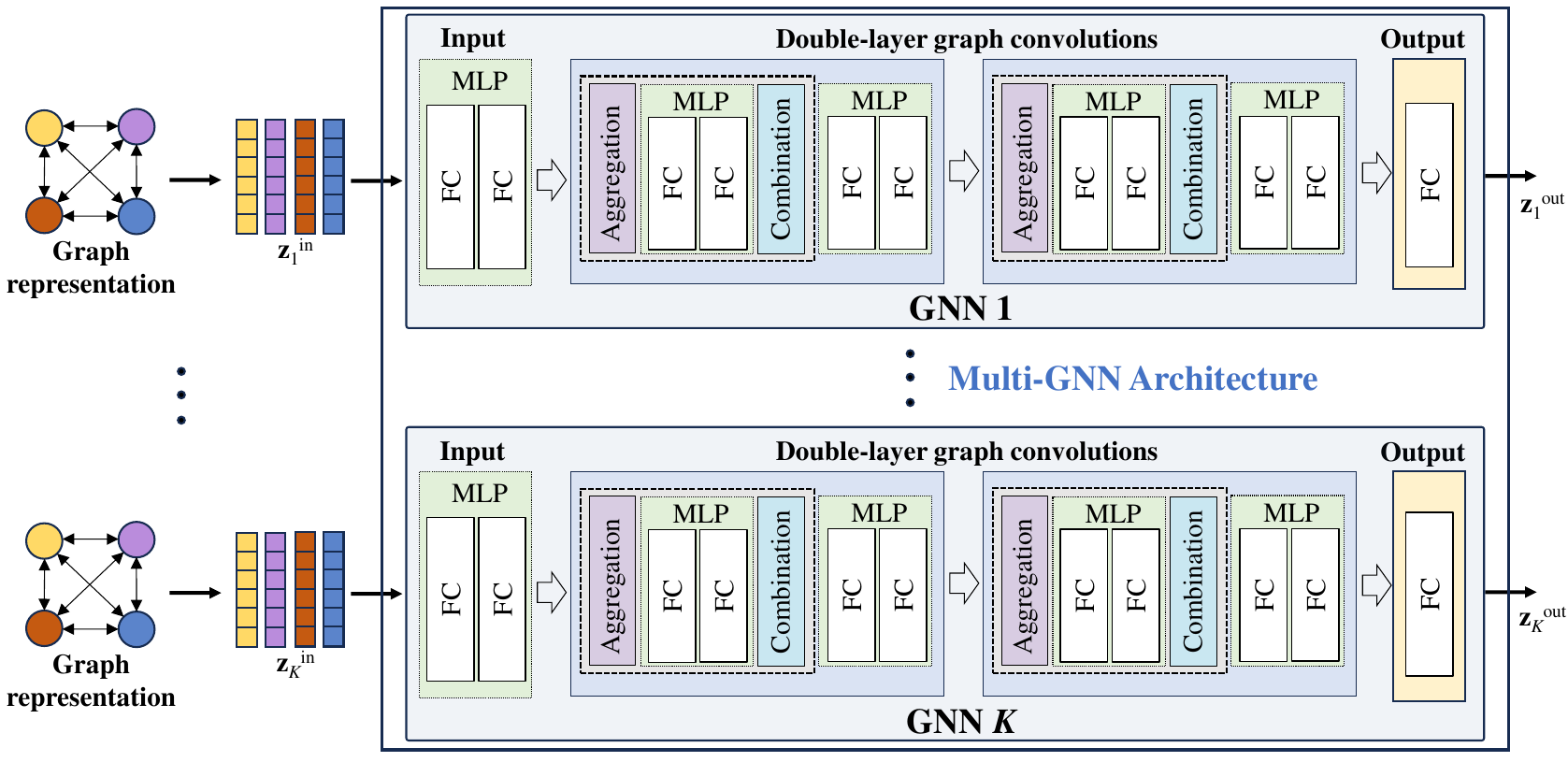}
\caption{An illustration of the multi-GNN architecture.}
\label{Fig3}
\end{figure*}
In the proposed multi-GNN architecture, all the GNNs are homogeneous and adhere to the same operational rules. For any single GNN, the neural network comprises a multi-layer perceptron (MLP) for initial input, double-layer graph convolutions and a fully-connected (FC) layer with post-processing for final output, as illustrated in Fig. \ref{Fig3}. Without loss of generality, we take the $k$-th GNN with $k \in \mathcal{K}$ as an example, where $\textbf{z}_{k}^{\text{in}}$, $\textbf{z}_{k}^{(d)}$ and $\textbf{z}_{k}^\text{out}$ are used to denote the initial input, the intermediate output through the $d$-th layer, and the final output representation vectors, respectively.
\subsubsection{MLP for Initial Input}
In the GNN, MLP is used as the initial layer. As mentioned in Section III-A, the input of GNN is a graph representation vector, which is derived from the corresponding local network consisting of one LEO satellite and all the UTs. Based on the system model, the channel vectors $\textbf{h}_{k,m}$ from the $k$-th LEO satellite to all the UTs are selected as the input representation vector that contains the feature of the UT nodes. Adopting the real representation of channel vectors, the input representation vector is given by
\begin{align}
\textbf{z}_{k}^{\text{in}} = [\text{Re}\{\textbf{h}_{k,m}\}, \text{Im}\{\textbf{h}_{k,m}\}], ~m \in \mathcal{M}. \nonumber
\end{align}
Since each LEO satellite is equipped with $N$ antennas, the number of the input nodes at the MLP layer is set to $2N$. Note that the channel vectors of different UTs are fed in parallel into the MLP layer. Through the MLP layer, the intermediate output representation vector $\textbf{z}_{k}^{(1)}$ is produced. Mathematically, the intermediate output representation vector is given by
\begin{align}
\textbf{z}_{k}^{(1)} = f_\text{MLP}(\textbf{z}_{k}^{\text{in}}), \nonumber
\end{align}
where $f_\text{MLP}(\cdot)$ denotes the operational function of the MLP layer consisting of two FC layers with the same activation function.
\subsubsection{Graph Convolutions}
In the GNN, double-layer graph convolutions are the core component, where the two graph convolutional layers have the same structure and operational rules. In each graph convolutional layer, there are two MLP layers. Here, we take the first graph convolutional layer as an example for illustration. When $\textbf{z}_{k}^{(1)}$ is fed into the first graph convolutional layer, it first passes through the first MLP layer, followed by an aggregation and combination process. Then, it passes through the second MLP layer, resulting in the output $\textbf{z}_{k}^{(2)}$. Mathematically, the update rule for any one UT node is given by
\begin{align}
z_{k,m}^{(2)} = f_\text{com} \left(z_{k,m}^{(1)}, f_\text{agg} ( \{ z_{k,m'}^{(1)} \}_{m' \in \mathcal{M} \setminus m} ) \right), \nonumber
\end{align}
where $z_{k,m}^{(2)}$ denotes the $m$-th element of $\textbf{z}_{k}^{(2)}$. $z_{k,m}^{(1)}$ and $z_{k,m'}^{(1)}$ denote the $m$-th element and the $m'$-th element of $\textbf{z}_{k}^{(1)}$, respectively. Note that $m$ is also the index of the $m$-th UT node of graph representation. $f_\text{com}$ and $f_\text{agg}$ represent the combination and aggregation functions of the graph convolutional layer, respectively.\par
Generally, the aggregation and combination functions have a huge effect on the scalability and generalization of a GNN. In our GNN, the aggregation function is given by
\begin{align}
f_\text{agg} ( \{ z_{k,m'}^{(1)} \}_{m' \in \mathcal{M} \setminus m} ) = \psi ( \{ f_\text{MLP} (z_{k,m'}^{(1)} )  \}_{m' \in \mathcal{M} \setminus m} ),  \nonumber
\end{align}
where $\psi(\cdot)$ is a function that is invariant to the permutation of the inputs. Here, the aggregation function is set to a nested function of MLP and element-wise max-pooling, aiming to extract and aggregate the feature information from a node's neighbours. Based on such a function, the interaction among the UTs is embedded into the output graph representation. For the  combination function $f_\text{com}$, the combination operation is implemented, followed by a MLP layer. Through the two graph convolutional layers in the GNN, the interaction among different UTs is learned, which contributes to inter-UT interference suppression. It is worth emphasizing that the GNN-based optimization method can be generalized to the case of any other number of UTs as all the channels of UTs are directly concatenated into a single tensor as input.
\subsubsection{Output}
In the GNN, the double-layer graph convolutions are followed by a FC layer, which has $2 \times 2N$ output neurons. Due to the power constraint and the real-valued output, normalization, power adjustment and real-to-complex conversion are also performed following the output of the FC layer. Aligned with the initial input channel tensor, each row of the final output tensor is a power-normalized optimized beamforming vector for a specific data stream. For instance, $\textbf{w}_{k,m}$ represents the beamforming vector for the $m$-th data stream from the $k$-th LEO satellite.
\subsection{Training and Deployment of Multi-GNN Model}
\begin{figure}
\centering
\includegraphics[width= 3.5 in]{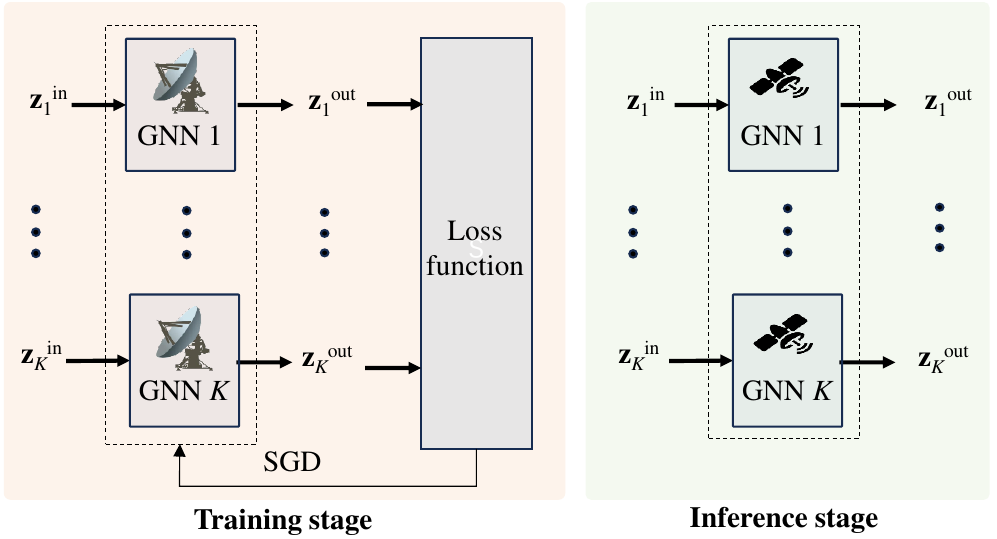}
\caption{The centralized training but distributed deployment of the multi-GNN model.}
\label{Fig4}
\end{figure}
In the considered LEO multi-satellite communications, it is appealing and realistic to implement centralized training but distributed deployment for the proposed multi-GNN model. As illustrated in Fig. \ref{Fig4}, the procedure of centralized training but distributed deployment is divided into three steps as follows.
\begin{enumerate}
\item The entire multi-GNN model is trained at the GW possessing abundant computing power and electricity. 
\item Each GNN from the trained model is separated and then deployed on the corresponding LEO satellite.
\item The weight and bias parameters of the trained multi-GNN model are delivered from the GW to each LEO satellite for the inference.
\end{enumerate}
\par In the training stage, the entire multi-GNN model is housed in the GW, granting it access to all the CSI involved in the multi-satellite system. The weight and bias parameters of the multi-GNN model are updated in an unsupervised manner, with the loss function defined as
\begin{align}
\mathcal{L} = -\frac{\sum_{t=1}^T \sum_{m = 1}^M \omega_{m} R_m^{(t)}}{T},  \nonumber
\end{align}
where $T$ denotes the size of training samples and $R_m^{(t)}$ is the achievable communication rate at the $m$-th UT for the $t$-th training sample. 
It is not hard to see that the objective function of optimization problem (P1) is inversely related to the loss function. As the loss function steadily decreases through stochastic gradient descent (SGD) during the learning process, the objective function gradually tends towards maximization. With the global CSI, the multi-GNN model can discern and learn the nuanced mechanisms governing both inter-satellite and intra-satellite interference. Since the training process is conducted offline, the computational complexity becomes less of a concern. \par
In the inference stage, the multi-GNN model is executed in a distributed manner. To be more precise, the entire model is decomposed into multiple GNN individuals, with only one individual GNN deployed on each LEO satellite. Consequently, each LEO satellite can directly complete the inference of its beamforming vectors using its GNN along with the associated local CSI. As the distributed deployment of GNN, the multi-GNN model has good scalability. It is worth mentioning that information exchange is necessary for the LEO satellites to implement synchronized coherent beamforming.
\begin{algorithm}[t]
\caption{Implementation of Graph Convolution Layer}
\begin{algorithmic}[1]
\State \textbf{Input}: $\textbf{x}_{i}^\text{in}$ with $i \in \mathcal{M}$
\State \textbf{Output}: $\textbf{x}_{i}^\text{out}$ with $i \in \mathcal{M}$
\For{$i$ = 1:$M$}
    \For{$j$ = 1:$M$}
        \If{$j \neq i$}
            \State $\textbf{x}_{j}^{(1)} = f_{\text{MLP1}} (\textbf{x}_{j}^\text{in})$
        \EndIf
    \EndFor
    \State $\textbf{x}_{i}^{(2)} = f_\text{agg} ( \{\textbf{x}_{i'}^{(1)}\}_{i' \in \mathcal{M} \setminus i} )$
    \State $\textbf{x}_{i}^{(3)} = f_\text{com} \left(\textbf{x}_{i}^\text{in}, \textbf{x}_{i}^{(2)} \right)$
    \State $\textbf{x}_{i}^\text{out} = f_{\text{MLP2}} (\textbf{x}_{i}^{(3)})$
\EndFor
\end{algorithmic}
\end{algorithm}
\subsection{Complexity Analysis}
For the GNN, the pseudocode about how to implement the graph convolution layer is presented in Algorithm 1.
Given the substantial impact of online inference latency on the operation of LEO multi-satellite communications, it is essential to analyze the computational complexity of forward propagation for the multi-GNN model. Here, we solely focus on a single GNN, as each GNN from the multi-GNN model implements the inference independently.
In each GNN, the initial layer is implemented by MLP. Hence, its computational complexity is given by $\mathcal{O}(2MNL_1 +ML_1L_2)$, where $L_1$ and $L_2$ denote the number of nodes in the hidden layer and the output layer, respectively. Additionally, an activation function is employed at each neuron of the MLP. In each GNN, two graph convolutional layers are utilized, each incorporating two MLP layers and their activation function. Their total computational complexity is approximately given by $\mathcal{O}( 2 M[(M-1)M(L_3L_4+L_4L_5) + M(L_6L_7+L_7L_8)] )$, where $L_3$, $L_4$ and $L_5$ denote the number of nodes in the input layer, the hidden layer and the output layer of the first MLP, respectively, and $L_6$, $L_7$ and $L_8$ represent those of the second MLP. In each GNN, the output layer is implemented by FC. Hence, its computational complexity is given by $\mathcal{O}(2ML_8 N)$. To summarize, the total computational complexity is approximately equal to the sum of the computational complexities of these individual components.
\section{FPGA-Based GNN Accelerator}
\begin{figure}
\centering
\includegraphics[width= 3.0 in]{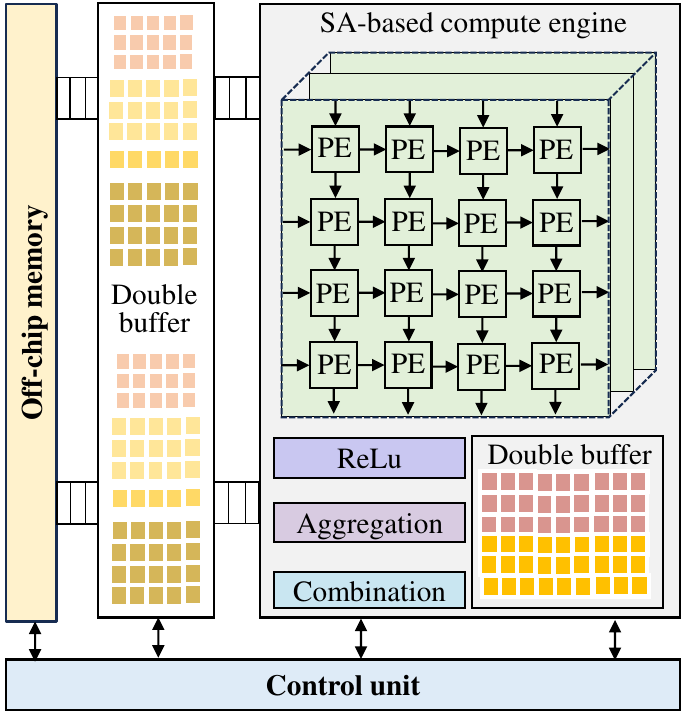}
\caption{The microarchitecture of the designed FPGA-based GNN accelerator.}
\label{Fig5}
\end{figure}
This section will design an FPGA-based accelerator to reduce the inference latency of GNN. Firstly, the microarchitecture of the FPGA-based accelerator is presented. Then, the communication between on-chip and off-chip memory is discussed. Lastly, the computation flow is introduced.
\subsection{Microarchitecture}
As illustrated in Fig. \ref{Fig5}. the designed FPGA-based GNN accelerator is comprised of three main components: computation engine, memory, and control unit. 
\subsubsection{Computation Engine}
In the adopted GNN, all the layers are related to FC, whose calculation depends on large matrix multiplication. Given this, multiple systolic arrays (SAs) are used as the core component of the computation engine. A SA is a specialized hardware architecture that consists of a grid of processing elements (PEs). Each PE is responsible for computing a partial product, accumulating the result, and passing data to its neighbors in a rhythmic, pulse-like manner. This structure allows for efficient parallel processing, making SAs particularly well-suited for tasks involving repetitive computations such as matrix operations. In addition to SAs, the computation engine also includes some other computing modules, including rectified linear unit (ReLU), aggregation and combination operations.
\subsubsection{Memory}
In terms of memory access, double buffering technology, specifically the ping-pong buffer technique, is employed to achieve parallel operations between data processing and data transfer, thereby enhancing system efficiency and reducing wait times. On the other hand, double buffering is also applied to store the output of one layer for use as the input for the next layer, contributing to layer fusion that combines multiple layers into a single group. This technique minimizes intermediate data transfers between on-chip and off-chip memory, thus lowering latency and reducing the energy consumption associated with excessive data movement.
\subsubsection{Control Unit}
The control unit employs a finite state machine (FSM) to manages computational counts and data access locations. By employing an FSM, the control module can systematically and efficiently manage both computational counts and data access locations, ensuring that operations are executed in the correct sequence and data is accessed precisely when needed. This structured approach reduces the complexity of control logic and enhances the overall performance of the system.
\subsection{Communication between On-Chip and Off-Chip Memory}
According to the roofline model~\cite{Zhang2015Optimizing}, computation and communication between on-chip and off-chip memory are two principal constraints for the inference latency of FPGA. As the considered GNN involves a substantial number of weight and bias parameters, the inference latency of the  FPGA-based GNN accelerator depends more on the I/O bandwidth accessing the off-chip memory. In other words, the implementation of the accelerator is memory-bound. Faced with such a situation, four key techniques are employed: quantization, loop tiling, double buffering, and layer fusion.\par
Quantization involves reducing the bit-width of data, typically from higher precision formats (like floating-point) to lower precision formats (like fixed-point or integer). This approach significantly reduces the latency of communication between on-chip and off-chip memory and improves computational efficiency on resource-constrained hardware platforms. Nevertheless, quantization may also result in potential accuracy loss. Overall, it is essential to balance precision and  efficiency during quantization.
Loop tiling aims to break down large loop bodies into smaller blocks. For the FPGA-based accelerator, only one tile can be processed at a time. Hence, all tiles are sequentially inputted into the FPGA-based accelerator for computation. This means that the accelerator can only begin computing the next tile once it has finished processing the current one. Through loop tiling, the loops are divided into on-chip and off-chip portions, which enhances on-chip data throughput.
For double buffering and layer fusion, please refer to Section IV-A.
\subsection{Computation Flow}
\begin{algorithm}[t]
\caption{Code Refactoring of Graph Convolution Layer}
\begin{algorithmic}[1]
\State \textbf{Input}: $\textbf{x}_{i}^\text{in}$ with $i \in \mathcal{M}$
\State \textbf{Output}: $\textbf{x}_{i}^\text{out}$ with $i \in \mathcal{M}$
\For{$i$ = 1:$M$}
    \State $\textbf{x}_{i}^{(1)} = f_{\text{MLP1}} (\textbf{x}_{i}^\text{in})$
\EndFor
\For{$i$ = 1:$M$}
    \State $\textbf{x}_{i}^{(2)} = f_\text{agg} ( \{\textbf{x}_{i'}^{(1)}\}_{i' \in \mathcal{M} \setminus i} )$ 
\EndFor
\For{$i$ = 1:$M$}
    \State $\textbf{x}_{i}^{(3)} = f_\text{com} \left(\textbf{x}_{i}^\text{in}, \textbf{x}_{i}^{(2)} \right)$
    \State $\textbf{x}_{i}^\text{out} = f_{\text{MLP2}} (\textbf{x}_{i}^{(3)})$
\EndFor
\end{algorithmic}
\end{algorithm}

The computation flow of FPGA-based GNN accelerator includes three steps. Firstly, the channel data of local communication network, the weight parameters and the bias parameters of the associated GNN model are fetched from off-chip memory to on-chip memory through double-buffers. Then, the computation is executed to output the beamforming data for the local communication network. During the computation, the code of double-layer graph convolutions in the GNN is refactored to reduce the count of invoking MLP layers by adapting to the size of designed SA, as shown in  Algorithm 2. Finally, the output is returned to off-chip memory. During this process, all the computations are performed on-chip while the external memory access is eliminated.
%
%
\section{Simulation and Experimental Results}
In this section, we firstly evaluate the achievable communication performance of the proposed GNN-based optimization method through numerical simulations. Then, we present the computing performance of the FPGA-based GNN accelerator.
\subsection{GNN-based Optimization Method}
\subsubsection{Simulation Setup}
\begin{table}[t]
\scriptsize
\begin{center}
\caption{Simulation Parameters.}\label{T1}~~\\
\begin{tabular}{|c|c|c|} 
\hline
\textbf{Notation}        &  \textbf{Description}                    & \textbf{Value}      \\ \hline
$d_0$           &  Height of LEO satellites       &  600 km           \\ \hline
$\lambda$       &  Carrier frequency              & 20 GHz          \\ \hline
$b_\text{max}$  &  Maximal antenna gain at each LEO satellite           & 52 dBi       \\ \hline
$P$             &  Maximum transmit power at each LEO satellite                    & 0 dBW        \\ \hline
$\sigma^2$             &  Noise variance at UTs               & -90 dBm          \\ \hline
$B$             &  Communication bandwidth        & 50 MHz           \\ \hline
$\varphi$             &  Beam angle between LEO satellites and UTs          &  $0.01^{\circ}$            \\ \hline   
$\varphi_\text{3dB}$             &  3-dB angle         & $ 0.4^{\circ}$        \\ \hline
$K$             &  Number of LEO satellites             & 4        \\ \hline
$N$           &  Number of antennas at each LEO satellite           & 4       \\ \hline
$M$      &  Number of UTs               & 4  \\ \hline
$\omega_{m}$   &  Weighting factor               & 1           \\ \hline
\end{tabular}
\end{center}
\end{table}
As listed in Table \ref{T1}, the involved simulation parameters for the considered satellite-terrestrial Space-MIMO communication system are set as follows. 
The height of all the LEO satellites is 600 km; the carrier frequency is $\lambda$ = 20 GHz; the maximal antenna gain at each LEO satellite is $b_\text{max}$ = 52 dBi; the maximum transmit power at each LEO satellite is 0 dBW; the noise variance at each UT is 90 dBm; the communication bandwidth is 50 MHz; the channel parameters of satellite-terrestrial links are  $(b_i, m_i, \Omega_i) = (0.063,2,8.97 \times 10^{-4}), \forall i\in \{p,e,s\}$; the beam angles between LEO satellites and UTs are $\varphi = 0.01^{\circ}$; the 3-dB angle is ${\varphi_\text{3dB}} = 0.4^{\circ}$; the number of LEO satellites is $K=2$; the number of antennas at each LEO satellite is $N=4$; the number of UTs is $M=4$; the weighting factors are $\omega_{m} = 1$. Some of these parameters may be used as variables in some simulation figures.
\par
In the simulations, several benchmark schemes are provided for comparison with the proposed GNN-based optimization scheme. These include minimum mean square error (MMSE), zero forcing (ZF) and maximum ratio transmission (MRT). To be more specific, 
\begin{itemize}
\item \texttt{GNN-Local}: This legend represents the proposed GNN-based optimization scheme, which requires only local CSI at each LEO satellite, as detailed in Section III.
\item \texttt{MMSE-Global}: This legend represents the MMSE scheme with global CSI, where all the LEO satellites work as a whole to implement the MMSE beamforming. In this scheme, we adopt an ideal assumption that the total transmit power constraint rather than individual power to show the performance upper bound of MMSE.
\item \texttt{MMSE-Local}: This legend represents the MMSE scheme with local CSI.
\item \texttt{ZF-Global}: This legend represents the ZF scheme with global CSI, where all the LEO satellites work as a whole to implement the ZF beamforming. In this scheme, we adopt an ideal assumption that the total transmit power constraint rather than individual power to show the performance upper bound of ZF.
\item \texttt{ZF-Local}: This legend represents the ZF scheme with local CSI, where the beamforming vectors at each LEO satellite are placed in the null space of their respective channel vectors.
\item \texttt{MRT-Local}: This legend represents the MRT scheme, where the beamforming vectors for data streams at each LEO satellite are parallel to its channel vectors, and the power is equally allocated among different data streams.
\end{itemize}
\subsubsection{GNN Setup}
\begin{table*}[t]
\scriptsize
\begin{center}
\caption{GNN Setup.}\label{T1I}~~\\
\begin{tabular}{|cc|cccc|cccc|c|}
\hline
\multicolumn{2}{|c|}{\textbf{MLP}}    & \multicolumn{4}{c|}{\textbf{GCN}}         & \multicolumn{4}{c|}{\textbf{GCN}}       & \multirow{3}{*}{\textbf{FC}} \\ \cline{1-10}
\multicolumn{1}{|c|}{\multirow{2}{*}{FC}} & \multirow{2}{*}{FC} & \multicolumn{2}{c|}{MLP}                                    & \multicolumn{2}{c|}{MLP}                & \multicolumn{2}{c|}{MLP}                                    & \multicolumn{2}{c|}{MLP}                &                              \\ \cline{3-10}
\multicolumn{1}{|c|}{}                    &                     & \multicolumn{1}{c|}{FC}      & \multicolumn{1}{c|}{FC}      & \multicolumn{1}{c|}{FC}       & FC      & \multicolumn{1}{c|}{FC}      & \multicolumn{1}{c|}{FC}      & \multicolumn{1}{c|}{FC}       & FC      &                              \\ \hline
\multicolumn{1}{|c|}{$2N$ * 1024}             & 1024 * 512            & \multicolumn{1}{c|}{512 * 512} & \multicolumn{1}{c|}{512 * 512} & \multicolumn{1}{c|}{1024 * 512} & 512 * 512 & \multicolumn{1}{c|}{512 * 512} & \multicolumn{1}{c|}{512 * 512} & \multicolumn{1}{c|}{1024 * 512} & 512 * 512 & 512 * $2N$                 \\ \hline
\end{tabular}
\end{center}
\end{table*}
In the adopted multi-GNN architecture, all the GNNs have the same structure. A single GNN consists of one MLP, two graph convolution layers, one FC layer, where the parameters of each layer are listed in Table \ref{T1I}. For all the FC layers throughout the entire GNN except for the final output layer, ReLU serves as the activation function. The SGD is carried out using the Adam optimizer with an initial learning rate set to 0.001. A learning rate decay strategy is employed by reducing the learning rate by a factor of 0.995 every 100 training steps. Each epoch utilizes 10,000 groups of randomly generated training samples, with a batch size of 200, resulting in 50 training steps per epoch. The testing dataset comprises 2000 samples. The training process concludes either upon reaching the predefined number of epochs or upon achieving convergence.
\subsubsection{Simulation Results}
\begin{figure}[t]
  \centering
  \subfigure[Convergence behavior]{\label{Fig6a}
            \includegraphics[width=1.65in]{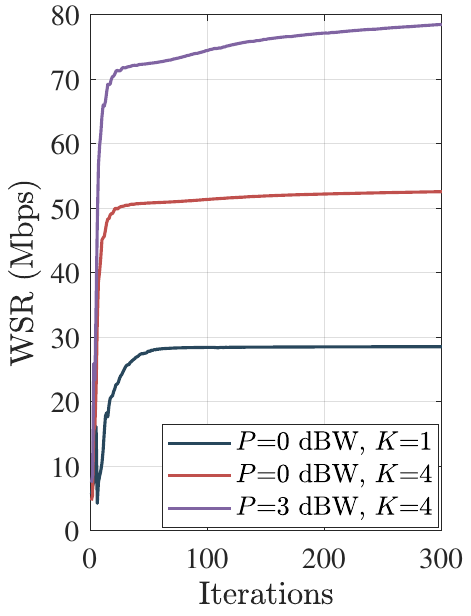}}
  \subfigure[Generalization]{\label{Fig6b}
            \includegraphics[width=1.6in]{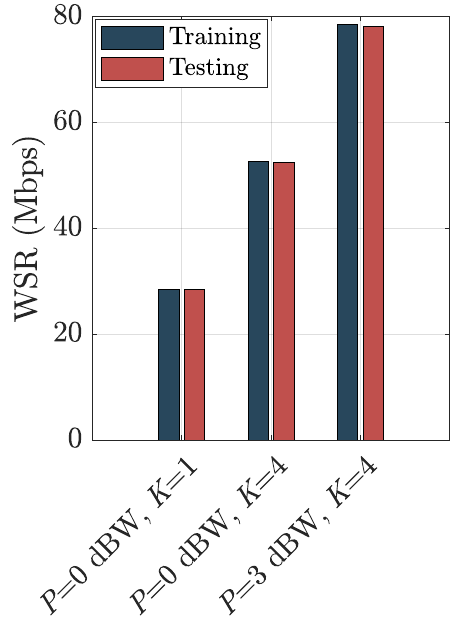}}
  \caption{Convergence and generalization behavior of the GNN-based optimization scheme under different settings of the parameters $P$ and $K$.}
  \label{Fig6}
\end{figure}
%
Fig. \ref{Fig6a} illustrates the convergence behavior of the GNN-based optimization scheme under different settings of the parameters $P$ and $K$. It is observed that the training of the GNN converges very quickly, stabilizing within 50 iterations, although it continues to grow very slowly afterward. Fig. \ref{Fig6b} evaluates the GNN model's performance and generalization capability by comparing the training results with the testing results. It is evident that the testing and training results are very close, confirming the model's robustness across various channel data.
\par
\begin{figure}
\centering
\includegraphics[width= 3.25 in]{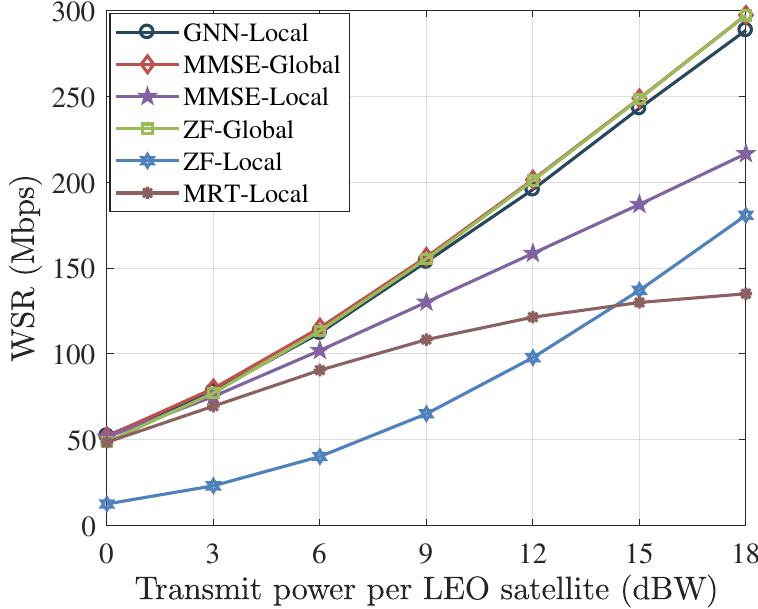}
\caption{The relationship between the WSR and the transmit power at each LEO satellite.}
\label{Fig7}
\end{figure}
Fig. \ref{Fig7} shows the relationship between the WSR and the transmit power at each LEO satellite. It is seen that an increase in the transmit power contributes to the enhancement of communication performance significantly. Compared to \texttt{MMSE-Local}, \texttt{ZF-Local}, and \texttt{MRT-Local}, \texttt{GNN-Local} achieves the highest WSR under identical CSI conditions. Additionally, the WSR of \texttt{GNN-Local} is slightly lower than that of \texttt{MMSE-Global} and \texttt{ZF-Global}. This is primarily because the schemes of \texttt{MMSE-Global} and \texttt{ZF-Global} are implemented with global CSI rather than local CSI. Furthermore, an ideal assumption of the total transmit power constraint is adopted. In other words, the schemes of \texttt{MMSE-Global} and \texttt{ZF-Global} have more relaxed constraints than \texttt{GNN-Local}. \par
\begin{figure}
\centering
\includegraphics[width= 3.25 in]{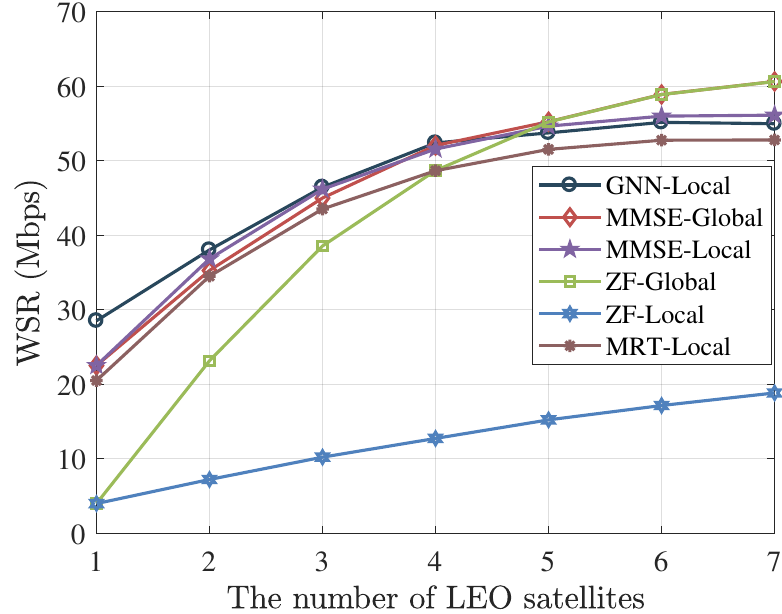}
\caption{The relationship between the WSR and the number of LEO satellites.}
\label{Fig8a}
\end{figure}

Fig. \ref{Fig8a} shows the relationship between the WSR and the number of LEO satellites. It is observed that the WSR of all schemes increases as the number of LEO satellites rises. This is because more LEO satellites offer a higher degree of spatial freedom. However, the growth rate slows down and the WSR tends to plateau, as the number of LEO satellites increases. It indicates that more LEO satellites aren't always better from the perspective of WSR. When the number of LEO satellites is fewer than four, the scheme of \texttt{GNN-Local} outperforms all other schemes. As the number of LEO satellites continues to increase, the performance of the scheme of \texttt{GNN-Local} declines relative to the schemes of \texttt{MMSE-Global} and \texttt{ZF-Global} but remains very close to \texttt{MMSE-Local}. \par
\begin{figure}
\centering
\includegraphics[width= 3.25 in]{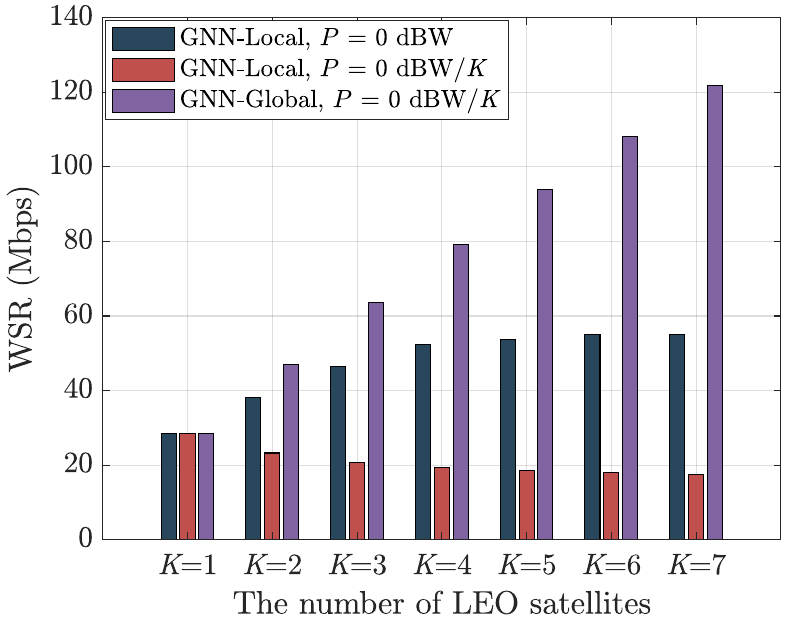}
\caption{The relationship between the WSR and the number of LEO satellites.}
\label{Fig8b}
\end{figure}

In Fig. \ref{Fig8a}, an increase in LEO satellites results in a linear increase in the total transmit power of system. To be more precise, the total transmit power  is given by 
\(P_{\text{total}} =  P \times K \). To mitigate the impact, more simulations are presented in Fig. \ref{Fig8b}, where two comparison schemes with the unchanged total transmit power \(P_{\text{total}}  =  P\) are shown. In the first comparison scheme, the transmit power of each LEO satellite is set to $P = 0~\text{dBW}/K$. In other words, the total transmit power of system is evenly distributed among all LEO satellites. It is observed that the achievable WSR decreases with the number of LEO satellites increasing. The reason is that the total transmit power of system is not flexibly allocated among LEO satellites, even though more LEO satellites can improve the degree of spatial freedom. In the other comparison scheme, namely \texttt{GNN-Global}, the antennas of all LEO satellites are combined into a single transmitter, allowing flexible power allocation. In this case, the achievable WSR increases rapidly as the number of LEO satellites grows. The performance gain is primarily attributed to the increase in the number of transmit antennas.\par
\subsection{FPGA-based GNN Accelerator}
This subsection will evaluate the performance of the FPGA-based GNN accelerator. In the experiment, Xilinx Virtex-7 XC7V690T FFG1761-3 FPGA is used as the hardware platform and the design is synthesized by Xilinx Vitis HLS 2022.2. \par
\begin{figure}
\centering
\includegraphics[width= 3.25 in]{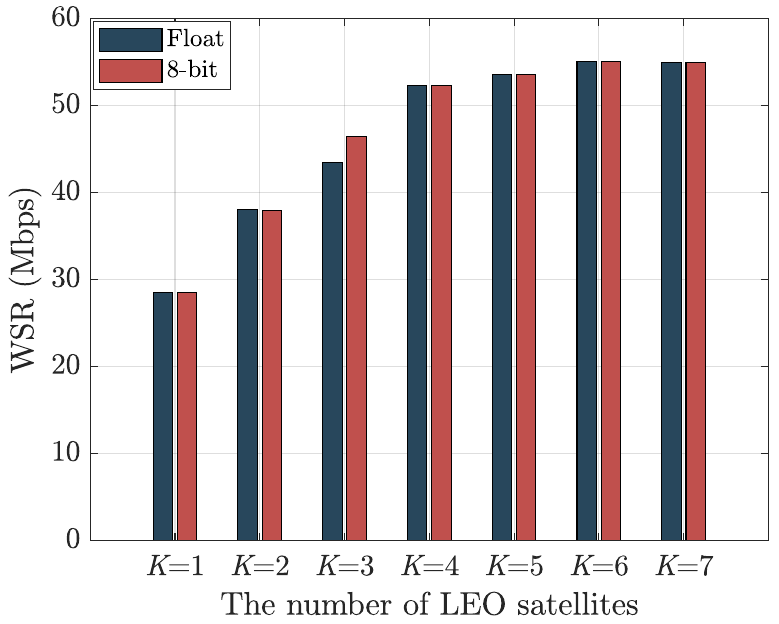}
\caption{The effect of fixed-point quantization on the achievable communication performance by the GNN-based optimization method.}
\label{Fig9}
\end{figure}
In the FPGA-based GNN accelerator, fixed-point arithmetic typically requires fewer hardware resources and consumes less power compared to floating-point arithmetic. Furthermore, the latency in communication between on-chip and off-chip memory is significantly reduced. However, fixed-point arithmetic often leads to some loss of accuracy for the inference of the trained neural network. Fig. \ref{Fig9} shows how the fixed-point quantization affects communication performance, where 8-bit fixed-point data representation is employed. Across varying numbers of GNNs in the adopted multi-GNN architecture, it is observed that the results of inference after quantization closely match the floating-point results.\par
As listed in Table \ref{T3}, the HLS tool’s FPGA resource consumption estimates are reported. In the hardware design, the GNN accelerator utilizes 64 bits of the off-chip data bandwidth, while the remainder is reserved for wireless communication and signal processing modules. Additionally, the latency ranges from 386,284 to 588,280 cycles, corresponding to 3.863 to 5.883 ms, when the target clock period is set to 10 ns. It is worth highlighting that the latency can be significantly reduced when the bit-width is increased and the quantization precision is lowered, given the implementation of the FPGA-based GNN accelerator is memory-bound. This can be confirmed by the example of 16-bit quantization, in which the latency ranges from 719,218 to 1,050,372 cycles, corresponding to 7.192 to 10.504 ms. These experimental results  demonstrate that the GNN-based optimization method can meet the latency requirements for most applications in the satellite-terrestrial Space-MIMO communication system.
\begin{table}[t]
\scriptsize
\begin{center}
\caption{Utilization Estimates.}\label{T3}~~\\
\begin{tabular}{|c|c|c|c|c|c|}
\hline
Name & BRAM\_18K & DSP & FF     & LUT    & URAM  \\ \cline{1-6}
Total  & 1472      & 128 & 100844 & 246646 & 0                    \\ \hline
Available  & 2940      & 3600 & 866400 & 433200 & 0                    \\ \hline
Utilization (\%)  & 50      & 3 & 11 & 56 & 0                    \\ \hline
\end{tabular}
\end{center}
\end{table}
%
%

%
\section{Conclusions}
This paper proposed a GNN-based Space-MIMO framework for D2C communications, named GSM. By developing a multi-GNN optimization method, distributed coordinated beamforming were efficiently implemented for LEO satellites. By customizing an FPGA-based GNN accelerator, the computation performance was significantly improved. The simulation results demonstrated that the proposed GNN scheme is superior to the benchmark ones including MRT, ZF and MMSE with local CSI but is slightly worse than ZF and MMSE with global CSI and global power allocation. Furthermore, the experimental results showed the inference latency of the FPGA-based accelerator is remarkably low, ranging from 3.863 to 5.883 ms when the target clock period is set to 10 ns and 8-bit fixed-point data representation is utilized.
%

%
%
%
%
%
%
\ifCLASSOPTIONcaptionsoff
  \newpage
\fi
\balance
%
%




\end{document}